\documentstyle[12pt,epsf,epsfig]{article}
\begin{document}

\date{\today}

\vspace{1cm}
\title{Charmed Hadron Production in $ \pi p $ and $ Kp $ Collisions 
and Leading Particle Effect by Constituent Quark-Diquark Cascade Model}
\vspace{1cm}

\author{\rm Tsutomu Tashiro$^1$,  Hujio Noda$^2$, \\
Shin-ichi Nakariki$^1$, Kazuo Ishii$^1$ 
and Kisei Kinoshita$^3$ \\
\\
$^1$Department of Simulation Physics, Faculty of Informatics \\
Okayama University of Science, Okayama 700, Japan \\
$^2$Department of Mathematical Science, Faculty of Science \\
Ibaraki University, Mito 310, Japan  \\
$^3$Physics Department, Faculty of Education \\
 Kagoshima University, Kagoshima 890, Japan}

\vspace{1cm}
\maketitle

\vspace{2cm}
\begin{abstract}
We investigate the meson, baryon and anti-baryon productions in 
$ \pi p $ and $ Kp $ collisions in terms of a constituent cascade 
model which includes charm flavour. The constituent diquark in an 
incident baryon has a tendency to get a large momentum fraction, 
resulting in considerable difference of subenergies for particle 
productions from quark-diquark and quark-antiquark chains in a 
meson-baryon collision.   As a consequence, the leading particle 
effects on $ D $ meson productions are expected to be quite different 
in $\pi^+ $ and $ \pi^- $ beams at present accelerator energies.\\
\\
pacs number: 13.85.Ni 
\end{abstract}

\vspace*{\fill}
\begin{center}
{\it Submitted to Zeitschrift f\"ur Physik C}
\end{center}

\vspace*{10mm}


\newpage
Experiments at CERN (WA82) and Fermilab (E769) measured charmed meson distributions and observed a strong asymmetry between leading and non-leading charmed meson productions in $ \pi$-nucleon collisions \cite{wa82,e769}.   The leading particle contains the same type of quark as one of the valence quarks in the incident hadron, while the non-leading one does not contain projectile valence quarks.   In perturbative QCD at leading order, the factorization theorem predicts that $ c $ and $ \bar{c} $ quarks are produced with the same distributions and then fragment independently.  Thus, the perturbative QCD model can not expect such a large asymmetry and contradicts with the experimental data \cite{qs}.  Even in the case of next to leading order, the predicted asymmetry is small \cite{nde}.  Alternatively, the final interaction of charm quark with spectator quarks may be responsible for the large asymmetry.  The Lund string fragmentation model implemented with this mechanism, PYTHIA Monte Carlo program, reproduces the trends of increasing asymmetry with Feynman $ x $, but the asymmetry is too large \cite{PYTHIA}.  Recently, Vogt and Brodsky took account of an intrinsic $ c \bar{c} $ in projectile hadron state and predicted an asymmetry from coalescence of charm quarks with the comoving spectator quarks, in agreement with the data \cite{vb}.  The charmed hadron production was also discussed in the framework of the dual string model \cite{Armesto}.  At the present stage, the asymmetry problem about the charmed hadron productions is in an open question. 

On the other hand, soft hadron productions have been extensively investigated in terms of constituent quark chain or constituent string models [8-15].  A constituent quark (diquark) is a quark-gluon cluster consisting of the valence quark (diquark), sea quarks and gluons.  Constituent quarks and diquarks play important roles in soft hadron interactions in contrast with parton quarks and gluons in hard interactions.  We have proposed a covariant quark-diquark cascade model and applied it to various types of hadron productions in hadron-hadron and hadron-nucleus collisions, successfully\cite{qdqA}.  General background, ideas and consequences of the cascade and chain type models are reviewed in \cite{kkst}. In our model, leading and non-leading particle effects on hadron spectra are naturally explained by the constituent cascade mechanism.  The momentum fractions of constituent quarks (diquarks) in projectile hadrons are characterized by parameters determined from Regge intercepts \cite{mnkt,cthkp}.  When two incident hadrons collide, they have a tendency to break up into fast and slow constituents.  Two cascade chains are exchanged between the constituents in the beam and target hadrons.  The leading particle effect results from the first cascade step of the energetic valence constituents.  It may be valuable to introduce charm flavour into our model and investigate the strong asymmetry between leading and non-leading charmed meson productions.

 In this paper, we calculate the non-charmed and charmed hadron productions in $ \pi p $ and $ Kp $ collisions in terms of the covariant quark-diquark cascade model.   The model is found to be able to explain the strong asymmetry between leading and non-leading charmed meson productions. Furthermore, it predicts different leading particle effects in $D$ meson productions between  $ \pi^+ $ and $ \pi^- $ beams.  In Sect.2, we formulate the model.  In Sect.3, the numerical analyses of non-charmed and charmed hadron spectra are given.  Conclusion and discussion are given in Sect.4.  

\section{Formulation of the model}
We consider a covariant cascade model \cite{km} in terms of constituent quark and diquark.  We assume that a baryon is composed of a constituent quark and a constituent diquark, and a meson is composed of a constituent quark and a constituent anti-quark.   Hereafter, we simply call them as a quark and a diquark.  We construct a Monte Carlo event generator to satisfy the energy-momentum conservation condition and apply it to hadron-hadron collisions.  We consider an inclusive reaction $ A+B \rightarrow C+X$.   In the centre of mass system of $ A $ and $ B$, the light-like variables of $ A $ and $ B $ are defined as follows:
\begin{eqnarray}
x_{0\pm}^A=\frac{E^A \pm p_{cm}}{\sqrt{s_0}},~~~x_{0\pm}^B=\frac{E^B \mp p_{cm}}{\sqrt{s_0}} ,
\end{eqnarray}
where $ \sqrt{s_0} $ is the centre of mass energy of the incident hadrons $A$ and $B$.

\subsection{Dissociation of incident hadrons and formation of the chains}
When the collision between $A$ and $B$ occurs, the incident hadrons break up into constituents with a probability $ (1 - P_{gl})$; otherwise they emit wee gluons with  $ P_{gl}$ followed by a quark-antiquark pair creation.  We assume four reaction types: a) non-diffractive dissociation, b),c) single-diffractive dissociation and d) double-diffractive dissociation types as shown in Fig. \ref{fgr:intrctn type}.   The probabilities of these types to occur are $ (1 - P_{gl})^2 , P_{gl} (1 - P_{gl}), P_{gl} (1 - P_{gl})$ and $ P_{gl}^2$, respectively.  Here we denote the quark-antiquark pair emitted from $ A $ ($ B $) via the wee gluons as $ M_A $ ( $ M_B $).  The probabilities of $ M_A $ ( $ M_B $) to be $u\bar{u}, d\bar{d}, s\bar{s} $ and $ c\bar{c} $  are denoted as $  P_{u\bar{u}},~ P_{d\bar{d}},~  P_{s\bar{s}} $ and $ P_{c\bar{c}}  $, respectively.  From the isospin invariance, we put $P_{u\bar{u}}=P_{d\bar{d}}$.   At very high energies, it is expected that 2$n$ cascade chains are generated in accordance with $n$ Pomeron exchange.  However we neglect here the processes with $ n \stackrel{>}{=} 2 $, since the incident energy to be considered is not so high.

  In$~$ the$~$ non-diffractive$~$ dissociation,$~$ both$~$ incident
\noindent
 hadrons $ A $ and $ B $ break up into two constituents. Two cascade chains are exchanged between $ A $ and $ B $.  Each cascade chain is developed from the two constituents, one of which is in the beam and the other in the target hadron. In the case of meson-baryon collisions, they are the quark-antiquark and the quark-diquark cascade chains as shown in Fig. 1a).   As shown in Ref. \cite{mnkt}, the distribution functions of the constituents in the incident hadron $ A $ composed of $ a $ and $ a' $ are described as
\begin{eqnarray}
 H_{a/A}(z) = H_{a'/A}(1-z) = \frac{z^{\beta_a-1}(1-z)^{\beta_{a'}-1}}{B(\beta_a,\beta_{a'})}. 
\label{eqn:Ha/A}
\end{eqnarray}
Then the light-like fractions of $ a $ and $ a' $ are $ x_+^a=x_{0+}^Az,~~~x_-^a=x_{0-}^AR,~~~~x_+^{a'}=x_{0+}^A-x_+^a,~~~x_-^{a'}=x_{0-}^A-x_-^a $, respectively.  Hereafter $ R $ denotes the uniform distribution in the interval $ 0 < R < 1 $.

   In the single diffractive dissociation, we assume that the diffractively dissociated hadron, say $ B,$ breaks up into two constituents and the other incident hadron emits a wee quark-antiquark pair $ M_A(q \bar{q}) $ and two cascade chains are exchanged between $ M_A $ and $ B $.  The $~$single-diffractive$~$ dissociation $~$mechanism $~$results$~$ in$~$ a

\noindent
diffractive peak of spectrum of the same kind of the incident particle $ A $ at $ x \approx 1 $.   We assume the distribution function
\begin{eqnarray}
  H_{M_A/A}(z)  = z^{\beta_{gl}-1}(1-z)^{\beta_{ld}-1}/B(\beta_{gl},\beta_{ld}) ,
\label{eqn:Hmaa}
\end{eqnarray}
with the light-like variables of $ M_A $ defined as,
\begin{eqnarray}
x_+^{M_A}=x_{0+}^Az,~~~x_-^{M_A}=x_{0-}^AR.
\label{eqn:Hmaa_2}
\end{eqnarray}
Then the leading particle $A$ and the diffractively dissociated hadron $B$ have the following momentum fractions:
\begin{eqnarray}
x_+^{A}=x_{0+}^A(1-z),~~x_-^{A}=m_A^2/(x_+^{A}~s_0),
\nonumber
\end{eqnarray}
\begin{eqnarray}
x_-^B=x_{0-}^B-(x_{-}^A-x_{0-}^A(1-R)),~~x_+^B=x_{0+}^B,
\label{eqn:Hmaa_2_leading}
\end{eqnarray}
where the mass shell condition is considered and transverse momentum of the leading particle is neglected.   In the centre of mass system of $ M_A $ and $ B $, we define the light-like variables of these hadrons and fix the light-like fractions of the projectile constituents from Eq.(\ref{eqn:Ha/A}) as in the non-diffractive dissociation.  The single-diffractive dissociation of $ A $ can be treated similarly, exchanging the role of $ A $ and $ B $.

In the double-diffractive dissociation, both the beam $A$ and target $B$ emit wee mesons $M_A$ and $M_B$ and then break into two constituents. The momentum fractions of $M_A$ and $M_B$ are determined by Eq. (\ref{eqn:Hmaa}) in the same manner as in the case of single-diffractive dissociation.  We have
\begin{eqnarray}
x_+^{M_A}=x_{0+}^A z,~~x_-^{M_A}=x_{0-}^A R,
\nonumber
\end{eqnarray}
\begin{eqnarray}
x_-^{M_B}=x_{0-}^B z',~~x_+^{M_B}=x_{0+}^B R'.
\label{eqn:Hmaa_3_leading}
\end{eqnarray} 
The$~$ momentum $~$fractions$~$ of $~$diffractively dissociated

\noindent 
hadrons $ A $ and $ B $ are
\begin{eqnarray}
x_+^{A} = x_{0+}^A(1-z),~~x_-^{A}=x_{0-}^A(1-R),
\nonumber
\end{eqnarray}
\begin{eqnarray}
x_-^{B} = x_{0-}^B(1-z'),~~x_+^{B}=x_{0+}^B(1-R').
\label{eqn:Hmaa_4_leading}
\end{eqnarray} 
Two cascade chains are exchanged between $ M_A $ and $ B $ and between $ A $ and $ M_B $ as shown in Fig. 1d.  Also the momentum fraction of the constituents are fixed by Eq.  (\ref{eqn:Ha/A}) in the centre of mass system of $ M_A $ and $ B $ ($ A $ and $ M_B $), as in the non-diffractive dissociation case.

The dynamical parameters $ \beta$'s in Eq. (\ref{eqn:Ha/A}), which determine the momentum sharings of the constituents in the incident hadrons, are related to the intercepts of the Regge trajectories as \cite{mnkt,cthkp}
\begin{eqnarray}
\beta_u=\beta_d=1-\alpha_{\rho-\omega}(0),~~ \beta_s=1-\alpha_\phi(0),
\nonumber
\end{eqnarray}
\begin{eqnarray}
\beta_c=1-\alpha_{J/\psi}(0) . 
\label{eqn:beta}
\end{eqnarray}
From previous analyses \cite{qdq,qdqA}, we determine the values for diquarks as
\begin{eqnarray}
\beta_{[ij]}=\gamma_{[~]}(\beta_i+\beta_j),~~\beta_{\{ij\}}=\gamma_{\{\}}(\beta_i+\beta_j),  
\label{eqn:betaij}
\end{eqnarray}
where [ij] and \{ij\} denote the flavour antisymmetric and symmetric 
diquarks, respectively.

The parameters $\gamma_{[~]}$ and $\gamma_{\{\}}$ are related to the proton structure and determined from the ratio $\pi^+/\pi^- $ in proton fragmentation. The parameters $\beta_{gl}$ and $\beta_{ld}$ in Eq. (\ref{eqn:Hmaa}) and $ P_{gl}$ are chosen to reproduce the diffractive peaks of $\pi^\pm p \rightarrow  \pi^\pm X$ at $ x \approx 1$ and  $\pi p \rightarrow  p X$ at $ x \approx -1$.  The ratio $ ~ P_{s\bar{s}}~/~P_{u\bar{u}} ~$ is fixed by the ratio $\pi^+/K^+$ in the proton fragmentation region. The probability $ P_{c\bar{c}} $ is determined from the $ D^\pm $ cross section in $\pi p$ collisions.

\subsection{Hadron productions from the cascade-chains}

Secondary hadrons are produced from the cascade chains exchanged between the two hadrons: (a)$ A $ and $ B $, (b) $ A $ and $ M_B $, (c) $ M_A $ and $ B $, (d)  $ A $ and $ M_B $, and $ M_A $ and $ B $.  We consider lower lying hadrons: pseudoscalar($PS$), vector($V$), tensor($T$) mesons, octet($O$) and decuplet($D$) baryons composed of $ u,d, $ and $ s $ flavours and the correspondings with charm flavour.  We assume the corresponding meson production probabilities in the cascade processes as $ P_{PS},  P_V $ and $ P_T $ ($P_T =1-P_{PS}-P_V $) and the corresponding baryon productions as $P_O $ and $ P_D$ ($ P_D=1-P_O$).  Octet and decuplet baryons are described as    
\begin{eqnarray}
|8 > = \cos\theta|q[q'q'']> + \sin\theta|q\{q'q''\}>,
\nonumber
\end{eqnarray}
\begin{eqnarray}
|10 > = |q\{q'q''\}> . 
\end{eqnarray} 

  We take account of the following cascade processes: 

\noindent
\begin{eqnarray}
q & \rightarrow & M(q \bar{q}')+q',~~B(q[q'q''])+\overline{[q'q'']} , \nonumber \\
& &  B(q\{q'q''\})+\overline{\{q'q''\}} , \nonumber \\  
\overline{[q'q'']} & \rightarrow & \overline{B}(\bar{q}\overline{[q'q'']})+q,~~M(q\bar{q}')+\overline{[qq'']} , \nonumber \\
& & M(q\bar{q}')+\overline{\{qq''\}}, \nonumber  \\ 
\overline{\{q'q''\}} & \rightarrow & \overline{B}(\bar{q}\overline{\{q'q''\}})+q,~~M(q\bar{q}')+\overline{[qq'']} , \nonumber \\
& & M(q\bar{q}')+\overline{\{qq''\}},
\end{eqnarray}

\noindent
where $ M(q\bar{q}'') $ is also produced with the same probability as that of $ M(q\bar{q}')$ from a diquark, and meson production probabilities from $ q,\overline{[q'q'']} $ and $ \overline{\{q'q''\}}$ are $ 1-\epsilon, \eta_{[~]} $ and $ \eta_{\{\}} $, respectively.  We assume equal probabilities of symmetric and anti-symmetric diquak productions from a diquark for $q \not\neq q''$. The pair creation probabilities $  P_{u\bar{u}},~ P_{d\bar{d}},~  P_{s\bar{s}} $ and $ P_{c\bar{c}}  $ are also assumed in the cascade process. The probabilities of $ [qq']\overline{[qq']}, \{qq'\}\overline{\{qq'\}} $ and $ \{qq\}\overline{\{qq\}} $ pair creations from a quark are chosen as $ P_{q\bar{q}}P_{q'\bar{q}'}, P_{q\bar{q}}P_{q'\bar{q}'}$ and ${P_{q\bar{q}}}^2$, respectively.  Then the emission probabilities of individual cascade processes are determined from the above probabilities.  For examples, the probabilities to produce $~~\pi^+, ~\rho^+,~a_2^+,...,~\Delta^{++}, ~p,\\
\Delta^+,..., \Xi_{cc}^{++}$ and $\Xi_{cc}^{*++}~$ from a $~u~$ quark are as follows: \\

\noindent
$(1~-~\epsilon)~P_{d\bar{d}}~P_{PS},~~
(1~-~\epsilon)~P_{d\bar{d}}~P_V,~~
(1~-~\epsilon)~P_{d\bar{d}}~P_T,..., \\
\epsilon~ P_{u\bar{u}}^{~~2} ,~~ 
\epsilon~ P_{u\bar{u}} ~P_{d\bar{d}}~(1 ~+~\frac{1}{3} ~P_O~\sin^2\theta / (~\frac{1}{3}~ P_O ~\sin^2\theta ~+ ~\frac{2}{3}~P_D)~),~~\\
\epsilon ~P_{u\bar{u}} ~P_{d\bar{d}}~ \frac{2}{3}~P_D /~(\frac{1}{3} ~P_O ~\sin^2 \theta ~+ ~\frac{2}{3} ~P_D ),...,\nonumber  \\ 
\epsilon ~P_{c\bar{c}}^{~~2}~\frac{2}{3}~P_O ~\sin^2 \theta/~(\frac{2}{3} ~P_O ~\sin^2 \theta ~+ ~\frac{1}{3} ~P_D~),~~ 
\epsilon ~P_{c\bar{c}}^{~~2}~\frac{1}{3}~P_D /~(\frac{2}{3} ~P_O ~\sin^2
\theta~+ ~\frac{1}{3} ~P_D),\nonumber \\ 
$

\noindent
where the factors 1/3 and 1/2 are flavour SU(4) factors.\\

We redefine the light-like fractions of the incident constituents in the rest frame of the cascade chain.  The momentum sharing of the cascade process $ q + \bar{q}' \rightarrow M(q\bar{q}'') + q''+ 
\bar{q}' $ from a $q $ with $ x_\pm^q $ and $ \bar{q}' $ with $ x_\pm^{\bar{q}'} $ takes place as follows \cite{qdqA,km}:  First, using the emission function
\begin{eqnarray}
 F_{q''q}(z) = z^{\gamma\beta_q-1}(1-z)^{\beta_q+\beta_{q''}-1} /B(\gamma\beta_q,\beta_q+\beta_{q''}) , 
\label{eqn:Fqq} 
\end{eqnarray}
we fix the lightlike fractions of $ q'' $ and $ M $ as $ x^{q''}_{+} = x^q_{+} z $ and $ x^{M}_{+} = x^q_{+} - x^{q''}_{+} $, respectively and put $ x^{q''}_{-} = x^q_{-} $.  Second, the transverse 
momentum of $ M $ is determined from the probability function
\begin{eqnarray}
   G(\mbox{\boldmath$p$}_{T}^2)=\frac{\sqrt{m}}{C}\exp(-\frac{C}{\sqrt{m}}\mbox{\boldmath$p$}_{T}^2)
\label{eqn:pT2} 
\end{eqnarray}
in $ p_{T}^2 $ space. Then, from the onshell condition, $ x^{M}_{-} $ is fixed as  $ x^{M}_{-} = (m_M^2 + {{\mbox{\boldmath$p$}}_{T}}^2)/x^M_{+}s'$, where $ \sqrt{s'} $ is the subenergy of the cascade chain. The transverse momentum of $ q'' $ is $ {\mbox{\boldmath$p$}}^{q''}_T = {\mbox{\boldmath$p$}}^q_T - {\mbox{\boldmath$p$}}_{T} $.  The lightlike fraction of $ \bar{q}' $ is decreased to $ \tilde{x}^{\bar{q}'}_{-} = x^{\bar{q}'}_{-} - x^{M}_{-}$.  If the energy of $\bar{q}'$ is enough to create another hadron, the cascade such as $ q'' + \bar{q}' \rightarrow q'' + \bar{q}''' + M(q'''\bar{q}') $ takes place in the opposite side.

The production probabilities $ P_{PS} $ and $P_V $ are fixed from $ \pi,~ \rho,~ a_2$ spectra. The probabilities $\epsilon$ and $ P_O$ are chosen so as to reproduce baryon productions in meson fragmentations.  The parameters $ \eta_{[~]} $ and $ \eta_{\{\}} $ are related to the spectra of the baryons which contain no projectile diquarks in the baryon beam fragmentation region.  The weight of antisymmetric diquark in the incident baryon $ \cos^2 \theta $ is chosen to get an appropriate $\pi^+/\pi^-$ ratio in the proton fragmentation region.  By introducing the dynamical parameter $ \gamma $ in Eq. (\ref{eqn:Fqq}), we can explain the $x$-distributions of various kind of hadrons by using the same parameters $\beta$'s in Eq. (\ref{eqn:Ha/A}). The parameter $C$ in Eq. (\ref{eqn:pT2}) is fixed from the experimental data on $p_T^2$ distributions.  These parameters are fixed so as to get an overall fitting to various hadron data with different incidents at different energies.

\subsection{Final step}
In the final step, the constituents from both sides terminate and recombine into one or two hadrons according to the processes: $
     q + \bar{q}' \rightarrow M(q\bar{q}'),~ 
     q + [q'q''] \rightarrow B(q[q'q'']), ~
     \{qq'\} + \overline{[q''q''']} \rightarrow B(q^*\{qq'\}) + \overline{B}(\overline{q^*}\overline{[q''q''']})$~ \\
and so on. The momenta of the recombined hadrons are the sum of those of the final constituents and are offshell.  The energy excess of the recombined hadron on $i$-th cascade chain is 
\begin{eqnarray}
\Delta E_{(i)}=E'_{(i)}-E_{(i)},
\nonumber
\end{eqnarray}
\begin{eqnarray}
E_{(i)}=\sqrt{ {\mbox{\boldmath$ p$}_{(i)}}^2 + m_{(i)}^2},~~~i=1,2,
\label{eqn:hh2} 
\end{eqnarray}
where $E'_{(i)}$ and $\mbox{\boldmath$ p$}_{(i)}$ are the sum of the energies and three momenta of the final constituents of $i$-th chain. There are different methods to put these particles on mass shell: (1) Multiplying the three momenta of all produced hadrons by a factor $\alpha$ so that the summation $ \sum_{\scriptstyle i=1}^{\scriptstyle n} \sqrt{(\alpha \mbox{\boldmath$ p$}_{i})^2 + m_{i}^2}$ would be equal to $ \sqrt{s_0}. $   (2) When $ m_{12}^2=(E'_{(1)}+E'_{(2)})^2-(\mbox{\boldmath$ p$}_{(1)}+\mbox{\boldmath$ p$}_{(2)})^2 >(m_{(1)}+m_{(2)})^2$, we can use two body decay $ m_{12} \rightarrow m_{(1)}+m_{(2)} $. Here we use the second method.  In order to minimize the effect of this procedure, however, we choose a pair of hadrons $j$ and $k$ which has the smallest value of
$
m_{jk}^2-(m_j+m_k)^2=(E_j+E_k+\Delta E_{(1)}+\Delta E_{(2)})^2-(\mbox{\boldmath$ p$}_{(1)}+\mbox{\boldmath$ p$}_{(2)})^2-(m_j+m_k)^2>0
$
among produced hadrons in the collision between $A$ and $B$.  In the two body decay, the directions of three momenta in $j + k$ system are conserved.  The energies and momenta of all directly produced hadrons are fixed.  Then directly produced resonances decay into stable particles.  For simplicity, their decays are assumed to be isotropic in the rest frame of the resonance.

\section{Comparison with data}
The newly introduced parameters concerning the charm flavour are the probability of $ c\bar{c} $ pair creation $ P_{c\bar{c}} $ and the parameter $ \beta_c$ fixing the momentum sharing of the charmed constituent in the cascade processes.  From the data on $ \pi^-p \rightarrow D^{\pm}X $ cross section \cite{ab}, we choose the parameters as  $ P_{c\bar{c}} = 0.00016 $ and $ P_{s\bar{s}} = 0.09984 $ . The dynamical parameter $ \beta_c $ is related to the Regge trajectory of $ J/\psi $.  Adopting the value of $ \alpha ' = 0.9 $, we take the value as $ \beta_{c} = 8.0 $.

We set other parameters related to the $u, d $ and $ s $ flavours to be nearly equal to the values used in the previous analyses \cite{qdq,qdqA}.  Namely, the pair creation probabilities of $ u\bar{u} $ and $ d\bar{d} $ are $  P_{u\bar{u}} = P_{d\bar{d}} = 0.45 $. Meson and baryon production ratios are chosen as $ P_{PS} : P_V : P_T = 0.4 : 0.4 : 0.2$ and $P_O : P_D = 0.33 : 0.67 $, respectively.   From $ \rho - \omega $ and $ \phi $ trajectories, $ \beta $ 's are chosen as $ \beta_u = \beta_d = 0.5 $ and $ \beta_s = 1.0$, respectively.  The dynamical parameters in Eq.(\ref{eqn:betaij}) and in Eq.(\ref{eqn:Fqq}) are chosen as $ \gamma_{[~]} = 1.5, \gamma_{\{\}} = 2.0 $ and $ \gamma = 1.75 $ .  The parameters $ P_{gl}, ~\beta_{gl} $ and $ \beta_{ld} $ are fixed from the diffractive peak of the projectile hadrons at $ |x| \approx 1. $   The probability of the incident hadron to emit a wee meson is set to be $ P_{gl} = 0.15 $ and the parameters $ \beta_{gl} $ and $ \beta_{ld} $  are chosen as $ \beta_{gl} = 0.1 $ and $ \beta_{ld} = 3.0. $   (Changing the parameters $ \beta_{gl} $ and $ ~\beta_{ld} $ do not alter the shape of the spectra so much.)  The mixing angle of symmetric and anti-symmetric diquarks in octet baryons is put as $ \cos^2 \theta=0.5 $.   The meson emission probabilities from quark and diquark are chosen as $ 1-\epsilon =1-0.07 $ and $ \eta_{[~]} = \eta_{\{\}} = 0.25 $, respectively. The $ p_T^2 $ distribution parameter in Eq.(\ref{eqn:pT2}) is chosen as $C=1.2$ in GeV unit.

\subsection{Non-charmed hadron spectra}

In Fig. \ref{fgr:pip250pt2}, the $ p_T^2 $ distribution of $\pi^-$ in $\pi^+ p$ collision is compared with the experimental data at $ p_L= 250 $ GeV/c \cite{Adamus}.  Method (1) suppresses the distribution at $ p_T^2 > 5$ (GeV/c)$^2$ as compared with method(2).  For $x$ distributions and asymmetries, the difference between methods (1) and (2) is very small and in the following we show only the results of method ~(2).   In Fig. \ref{fgr:pip3_7_100 to ps}, we show the results of $ \pi^\pm $ and $ K^\pm $ distributions in $ \pi^+p $ collisions at $ p_L=3.7 $ GeV/c \cite{Sheph} and 100 GeV/c \cite{Brenn}.  We note that the four momentum is conserved by our covariant formalism.  The $ s $-dependence as well as the $ x $ behaviour of the spectra are well reproduced. The kinematical constraints on particle productions reduce the spectra of $ \pi^- $ and $ K^- $ at $ p_L=3.7 $ GeV/c relatively as compared with those at $ p_L = 100 $ GeV/c.  The enhancement of the $ \pi^+ $ spectrum at $ x = 1 $ is due to the diffractive target dissociation type mechanism. The leading particles have harder $x$ distributions than the non-leading particles. From the choice of the dynamical parameters $ \beta $'s in Eqs.(\ref{eqn:beta}) and (\ref{eqn:betaij}), the incident $ u $ and $ d $ valence quarks in proton have softer distributions than the $ u $ and  $ \bar{d} $ valence quarks in $ \pi^+ $ .  Therefore the spectra of leading mesons in the proton fragmentation are softer than those in the $ \pi^+ $ fragmentation.  Even the non-leading $ K^- $ spectrum has very similar tendency, reflecting the initial configuration of the constituents of the chain.    In Fig. \ref{fgr:pdelta}, we show the result of proton distribution and compare with the data \cite{na22baryon}.  The peak of the proton distributions at $ x \approx -1 $ is also attributed to the single-diffractive dissociation type mechanism.

The $ x $ distributions of $ \rho^0 $ and $ K_s^0 $ in $ \pi^+ p$ and $ K^{*+} $ and  $ K_s^0 $  in   $ K^+ p $ collisions at $ p_L = 250 $ GeV/c \cite{na22} are shown in Figs. \ref{fgr:kpl250 to ps} a and b, respectively. The peaks of $ K_s^0 $ spectra at $ x \approx 0 $ are steeper than those of vector mesons in $ \pi^+ p $ and $ K^+ p $ collisions, reflecting resonance effects on $PS$ mesons.  The spectrum of $ K_s^0 $ in $ K^+ p$ interaction at $ x > 0 $ decreases more slowly with $ x $ than that of  $ K_s^0 $ in $ \pi^+ p$ interaction. Here we see  the effect of the constituent distributions i.e. the $ \bar{s} $ in $ K^+ $ is energetic as compared with the constituents $ u $ and $ \bar{d} $ in $ \pi^+ $ . The agreement with the data is fairly good.

\subsection{Charmed hadron spectra}
   Next we show the result of charmed meson productions. In Fig. \ref{fgr:dldnl}, the leading ($ D_{ld} $) and non-leading ( $ D_{nl} $) $ D $ meson productions with respect to the incident $\pi^-$ in $ \pi^- p $ collisions at $ p_L = 360 $ GeV/c are compared with the data \cite{ab}.  Here we mean $ D_{ld} $ and $ D_{nl} $  as  $ D_{ld} = D^- + (D^{*-} \rightarrow \overline{D}^0) $ and  $ D_{nl} = D^+ + ( D^{*+} \rightarrow D^0)$, respectively. From these data we fixed the value of $ P_{c\bar{c}} $.

We calculate the asymmetry $ A $ which is defined as
\begin{eqnarray}
A_{\pi^\pm} = \pm \frac{D^+ + D^{*+} - D^- - D^{*-}}{D^+ + D^{*+} + D^- + D^{*-}}
\end{eqnarray}
for $ \pi^\pm p $ collisions and compare with the data \cite{wa82,e769} in Fig. \ref{fgr:Dasymm}. The experiment WA82 measured the asymmetry by 340 GeV/c $ \pi^- $ beams and E769 used a mixed beam ($75\% \pi^-$ and $25\%  \pi^+ $) at 250 GeV/c.  Fig. \ref{fgr:Dasymm}b) shows our calculated $ p_T^2 $ dependence of $ A_{\pi^\pm} $ in the region $ 0.1 < x < 0.7 $ with the data of E769.   Our model predicts rather large positive values for $ \pi^\pm $ beams with $ A_{\pi^+} > A_{\pi^-}$.  The discrepancies between the prediction and the experimental data may be attributed to the neglect of multi-chain type mechanism and/or inadequate $p_T^2$ distribution. \\

In $\pi^- p$ interaction, the leading particle  $D^-$ is produced on the chain between the valence $d $ quark in the beam and the valence diquark in the target.  While in $\pi^+ p$ interaction, the leading particle $D^+$ is produced on the chain between the valence $\bar{d} $ quark in the $ \pi^+ $ beam and the valence quark in the target.  From Eqs.(4) and (5), the proton target has a tendency to break into an energetic valence diquark and a wee valence quark.  The former chain developed from $d$ quark and diquark in $\pi^- p$ collision is more energetic than the latter chain.   The leading $ D^-$ meson produced in the first cascade process is energetic.  Since the most energetic non-leading $D^+$ is produced after $ D^-$ production, $D^+$ is less energetic in our model.  Therefore the resulting asymmetry is large at $ x \approx 1$.  In the case in which a baryon is produced from the valence diquark in the first cascade process, the momentum of the valence $d$ quark is reduced.  Furthermore, when the valence $d$ quark has a small momentum, the total momentum is shifted to the diquark side.  Then the leading spectrum of $D^-$ on the quark-diquark chain tends to have a small momentum and the asymmetry is reduced at $ x < 0.7 $. On the other hand, the $ \bar{d} $ and the quark chain in the $ \pi^+ p$ interaction is less energetic than the quark-diquark chain.  So the leading $D^+$ production with a large momentum in the first cascade process is suppressed resulting a small asymmetry $A_{\pi^+}$ at $ x \approx 1$.  The momentum reduction of the $\bar{d}$ quark is small in the case when a meson is produced in the first cascade step from the valence quark. Therefore the asymmetry $ A_{\pi^+} $ is not so suppressed as compared with $ A_{\pi^-} $ at $ x < 0.7 $.

In the diffractive dissociation type mechanisms, the chain belongs almost to either the beam or the target fragmentation and particle productions have no  $\pi^\pm$ beam dependencies.   With the four types of interaction mechanisms the resulting asymmetry in $\pi^- p$ collision increases with $x$ and has a dip structure in the vicinity of $ x \approx 0.5$.  The asymmetry in $\pi^+ p$ has a peak at $x\approx 0.6$ and smaller than the one in $\pi^-$ beam at $ 0.8 < x$.

In Fig. \ref{fgr:pi&K to lamb} the $ \Lambda $ and $ \Lambda_c $ distributions computed in our model are shown with the $ \Lambda $ spectrum \cite{na22baryon}.  The resulting spectrum of $ \Lambda_c $ at $ x < 0 $ has a steeper peak than that of $ \Lambda $. This is attributed to the emission function with the large value of $ \beta_c $.   Leading baryon (which contains the same kind of diquark as the incident target) spectra have a peak at $ x < 0 $ due to the valence diquark effect. It is also noted that the non leading baryon spectra at $ x < 0 $ are also affected by the valence diquark in the target.  The shapes of the non-leading baryon spectra are softer than those of leading baryons, but are harder than those of antibaryons  and mesons in $ p $ fragmentation.  This is because of the effect of the processes $ qq \rightarrow M + qq'$.  In Fig. \ref{fgr:pi&K to lamb_a} we show our calculations of the $ \overline{\Lambda} $ and $ \overline{\Lambda_c} $ distributions and compare with the measured $ \overline{\Lambda}$ data \cite{na22baryon}. The $ \overline{\Lambda} $ spectrum in $ K^+ $ beam fragmentation shows a strong leading particle effect as compared with that in $ \pi^+ $ fragmentation.  This is because the valence $ \bar{s} $ quark  in the $ K^+ $ beam is harder than the valence $ \bar{d} $ quark in $ \pi^+ $ beam.

\section{Conclusion and discussion}
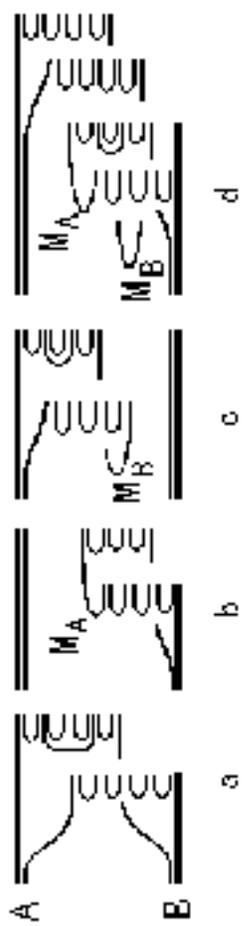
\begin{figure}
\caption{The reaction mechanisms in $AB$ collision:{\bf a}
 Non-diffractive dissociation type, {\bf b,c} Single-diffractive
 dissociation type and {\bf d} double-diffractive dissociation type mechanisms.}
\label{fgr:intrctn type}
\end{figure}

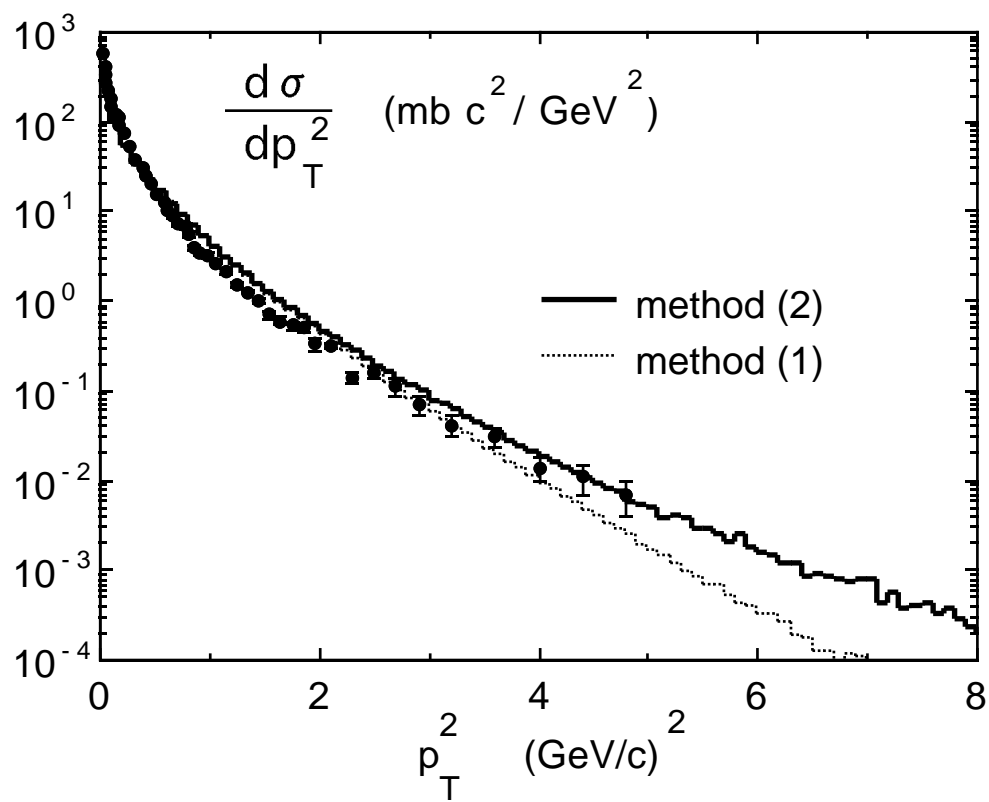
\begin{figure}
\caption{Calculated values for $ p_T^2 $ distributions of $ \pi^-$ in $ \pi^+p $ collisions at $ p_L=250 $ GeV/c. Experimental data are taken from [20]. }
   \label{fgr:pip250pt2}
\end{figure}

The covariant quark-diquark cascade model with charm flavour has been investigated.  The model can successfully explain the non-charmed hadron spectra.   The mass effects on the spectra are well explained by the flavour SU(3) breaking parameters $ P_{j\bar{j}} $ and the momentum sharing parameters $ \beta_j $.  As a result of the different hardness of the valence constituents in the incident hadrons, the leading particle effects on hadron productions are vastly different from each other. The order of the hardness of hadron spectra characterized in view of the constituent quark and diquark cascade is as follows: in decreasing order are $ p \rightarrow B, M \rightarrow M, p \rightarrow M, M \rightarrow B(\bar{B}$) and $ p \rightarrow \bar{B} $.  The detailed features of the spectra are described by the flavour dependencies of $ P_{j\bar{j}} $ and $ \beta_j$, and  by reaction mechanisms in Fig. \ref{fgr:intrctn type}.

   We have investigated charmed hadron spectra. The dynamical parameter $\beta_c$ determined by Eq.(\ref{eqn:beta}) and $P_{c\bar{c}}$ well explain charmed hadron productions.  It is noted that the dual-Regge ansatz for constituent quark distributions \cite{mnkt} appear to work even for charmed quark.  The momentum sharing functions in cascade processes and the beam dissociation functions are characterized by the same dynamical parameters determined from Regge intercepts.  Energy conservation condition with the constant values of $P_{j\bar{j}}$ well explains various hadron production ratios in the energy range 3.7 GeV/c $ \leq p_L \leq$ 360 GeV/c.

In our model, the asymmetry between leading and non-leading $ D^\pm $ meson productions depends on the incident beams.   The asymmetries $ A_{\pi^\pm} $ in Fig. \ref{fgr:Dasymm} reflect the distributions of $ \bar{d} $ in $ \pi^+ $ and $ d $ in $ \pi^- $ beam.  Although the distributions of the constituents in the incident $ \pi^{\pm} $ are the same, we have different asymmetries between $ \pi^+ $ and $ \pi^-\ $ beams.  In the single and double-diffractive dissociation, particle productions in the beam fragmentation are separated from those in the target fragmentation region.  However in the non-diffractive dissociation, the particle productions are mixed up with each other.  We have different cascade chain properties between the quark-antiquark and the  quark-diquark chain and have different leading particle effects between  $ \pi^+ $ and $ \pi^- $ beams.  Therefore the weight of the non-diffractive dissociation mechanism $ (1-P_{gl})^2 $ is also related to the the asymmetries $ A $ with respect to $\pi^\pm $ beams.

 Although it is interesting to apply the model to hadron productions containing bottom quark, their cross sections in soft interactions  are too small to discuss seriously at the present stage.  So we disregard the possibility of the hadrons containing bottom quark as a valence constituent.\\

\vspace{1cm}
Acknowledgement\\

We are grateful to Dr. H. Kohda and Dr. A. Iyono for helpful discussions. Part of this work was supported by the Grant-in-Aide for Scientific Research ((C)(2)(No.08640347)) of the ministry of Education, Science and Culture of Japan.

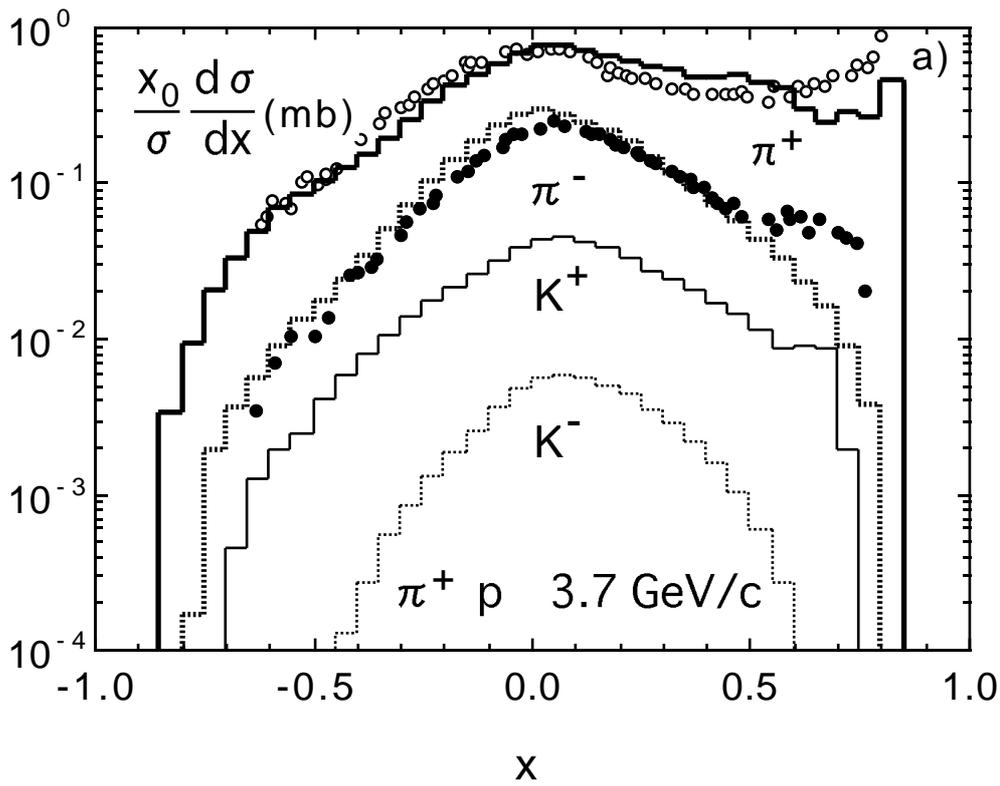
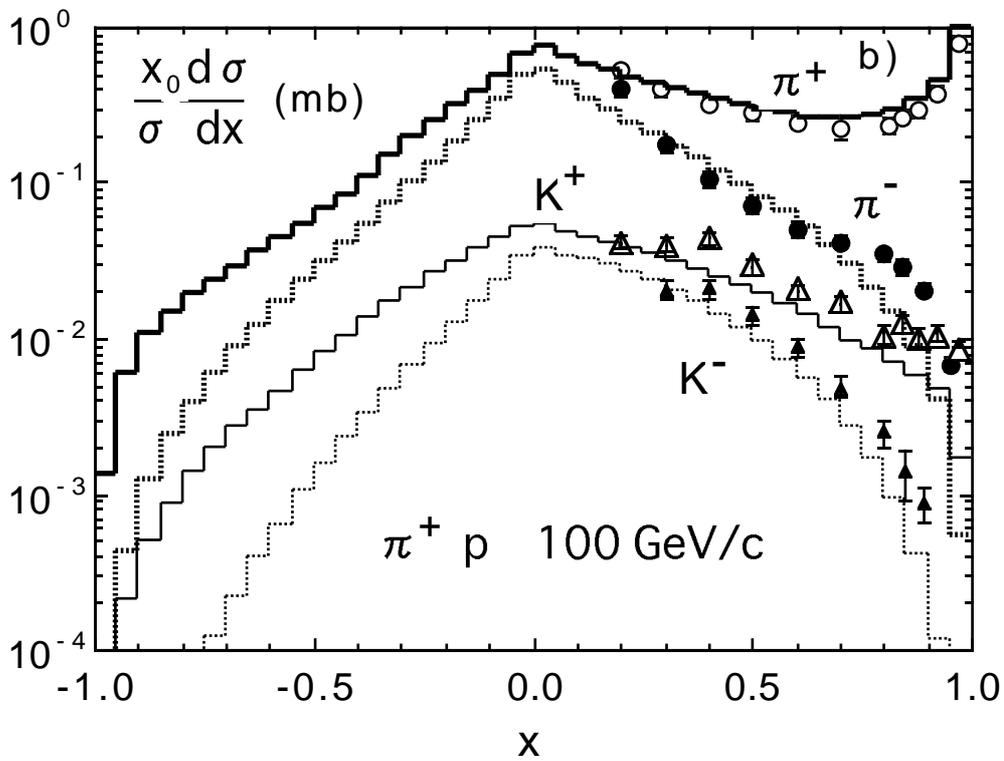
\begin{figure}
   \caption{Invariant $ x $ distributions of $ \pi^\pm$ and $K^\pm $  in $ \pi^+p $ collisions at {\bf a} $ p_L=3.7$ GeV/c  and {\bf b} $ p_L=100 $ GeV/c. Experimental data are taken from [21] and [22]. }
   \label{fgr:pip3_7_100 to ps}
\end{figure}
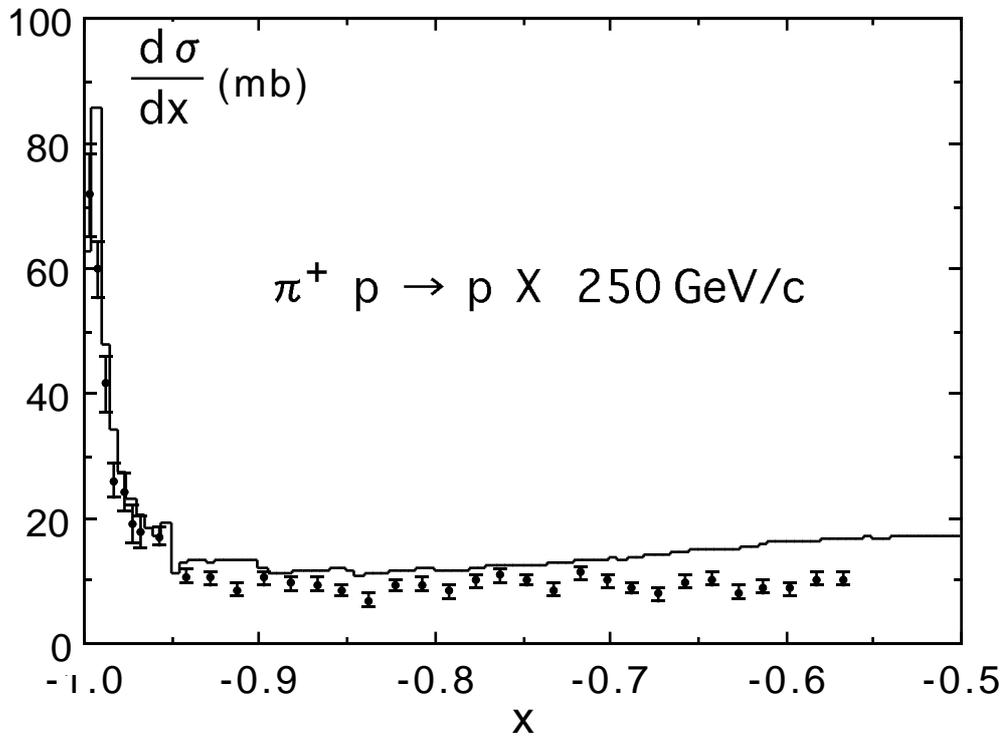
\begin{figure}
   \caption{Calculated values for $ x $ distributions of $ p $ in $\pi^+p$ collisions at $  p_L=250 $ GeV/c.   Experimental data are taken from [23] .}
   \label{fgr:pdelta}
\end{figure}

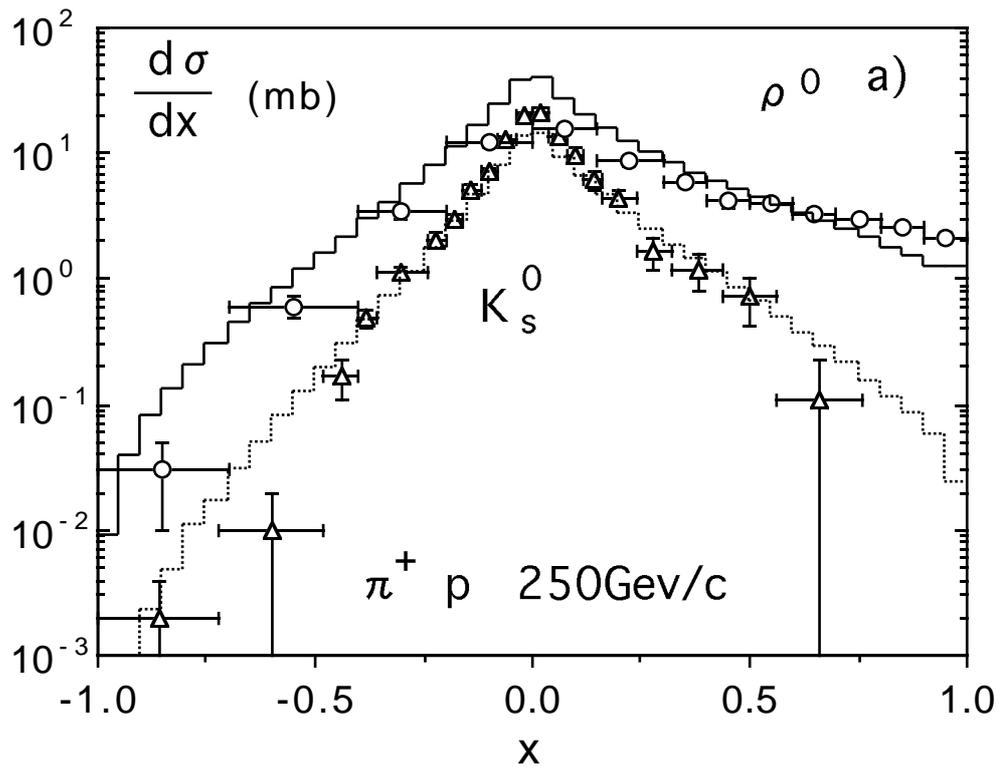
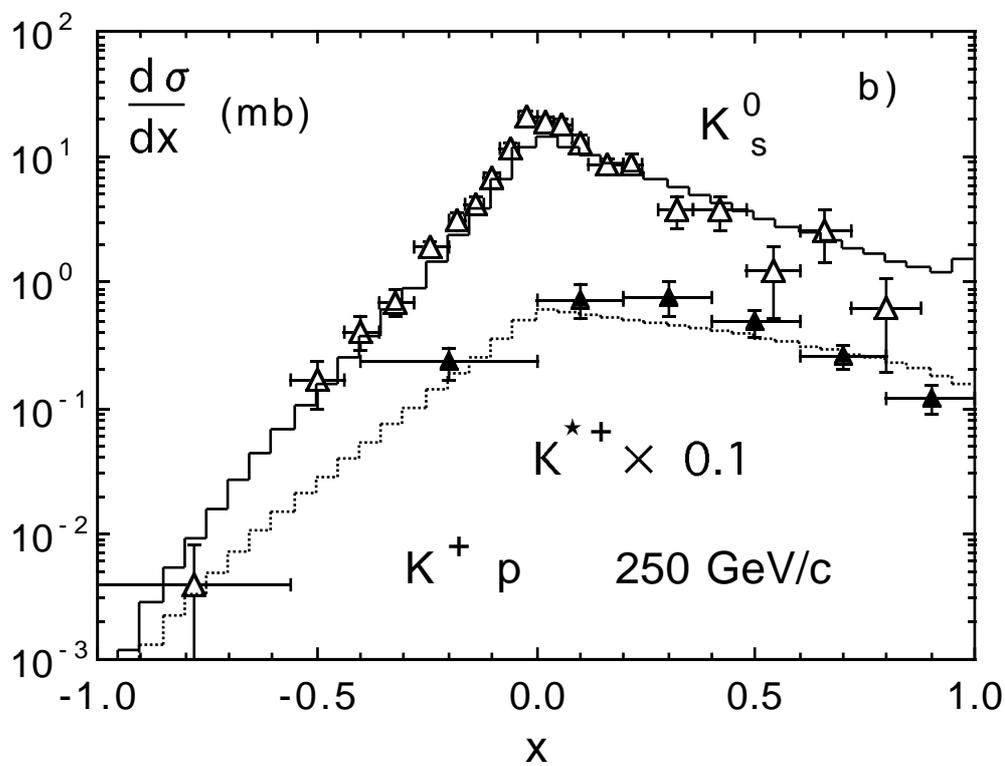
\begin{figure}
   \caption{Feynman  $ x $ distributions of {\bf a} $ \rho^0 $ and $ K_S^0 $ in $ \pi^+p $ and {\bf b} $ K_S^0 $ and $ K^{*+} $ in $ K^+p $ collisions at $ p_L=250 $ GeV/c.  Experimental data are taken from [24].}
   \label{fgr:kpl250 to ps}
\end{figure}

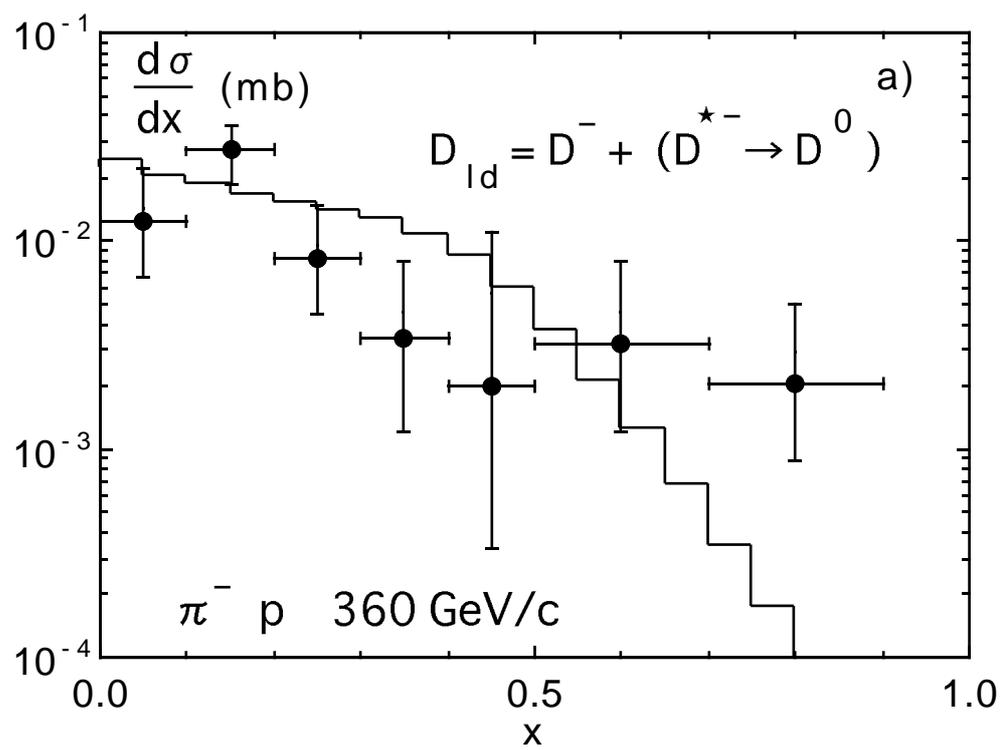
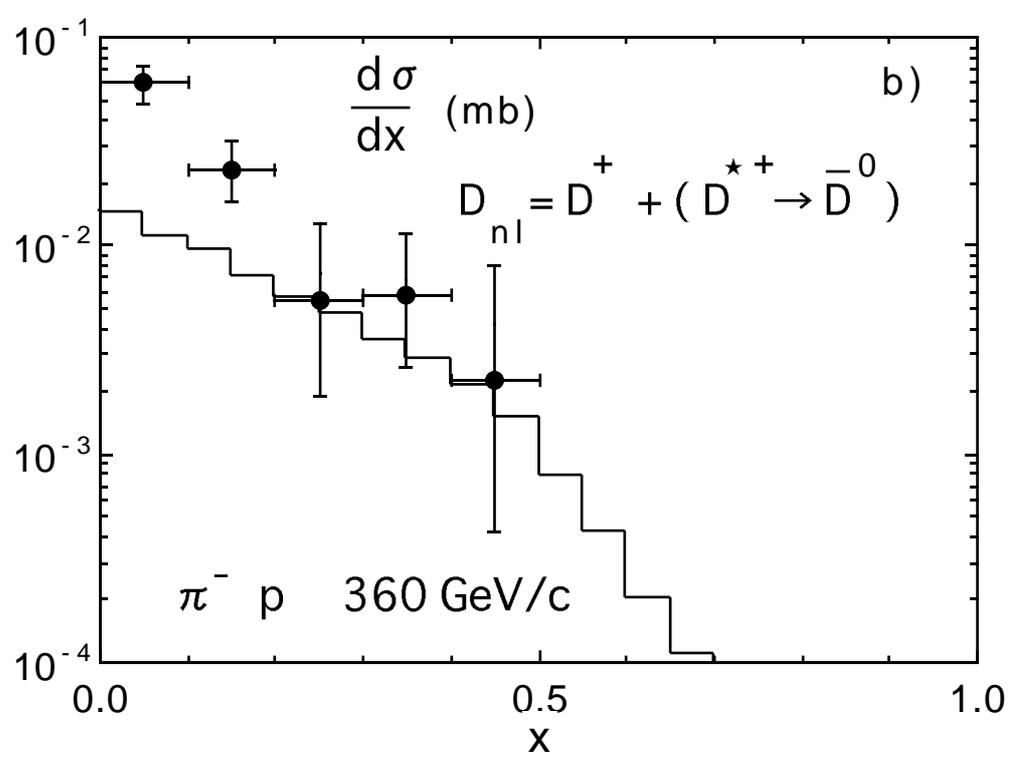
\begin{figure}
  \caption{Calculated values for $ x $ distributions of {\bf a} $ D_{ld}= D^- + (D^{*-} \rightarrow \overline{D}^0) $ and {\bf b} $ D_{nl}= D^+ + ( D^{*+} \rightarrow D^0) $ in $ \pi^- p $ collisions at $ p_L=360 $ GeV/c.  Experimental data are taken from [19]. } 
   \label{fgr:dldnl}
\end{figure}

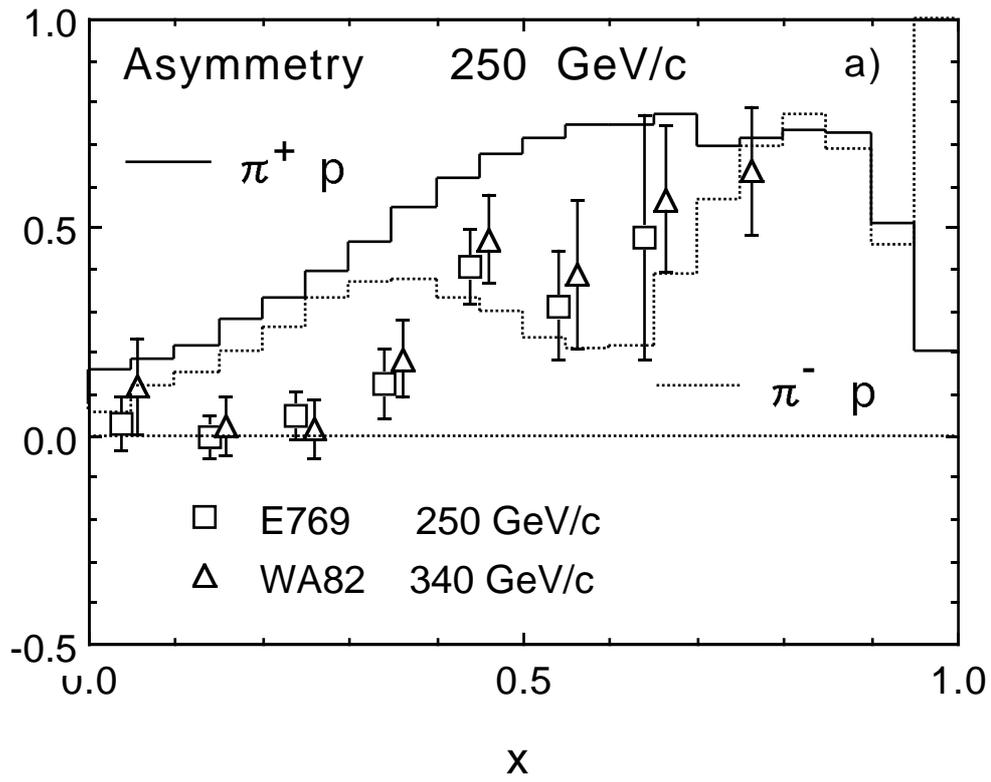
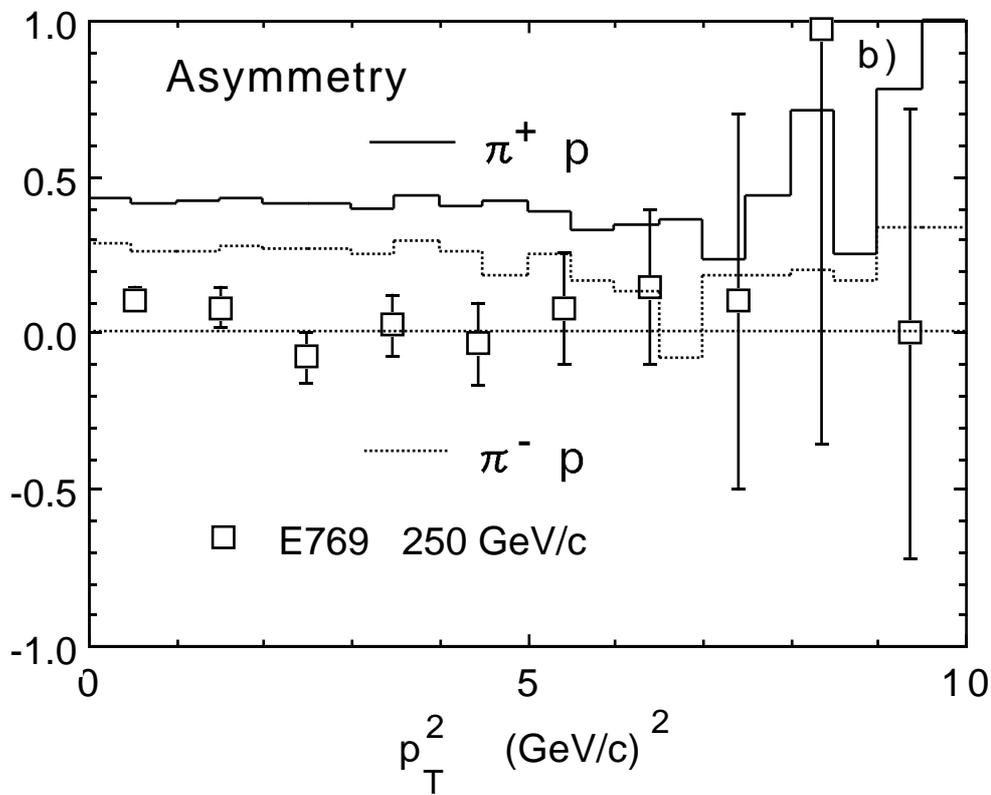
 \begin{figure}
  \caption{Asymmetries of $ D^\pm $ in $ \pi^\pm p $ collisions: {\bf a} vs. $ x $ and {\bf b} vs. $ p_T^2 $  at $ 0.1 < x < 0.7 $.  Experimental data are taken from [1] and [2].}
   \label{fgr:Dasymm}
\end{figure}

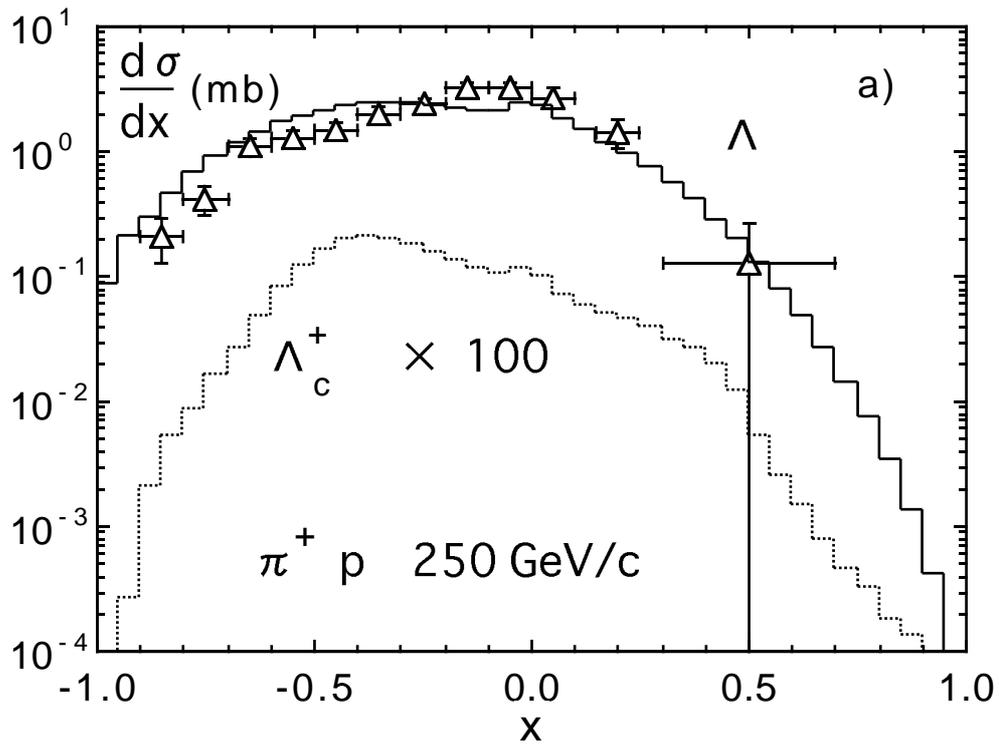
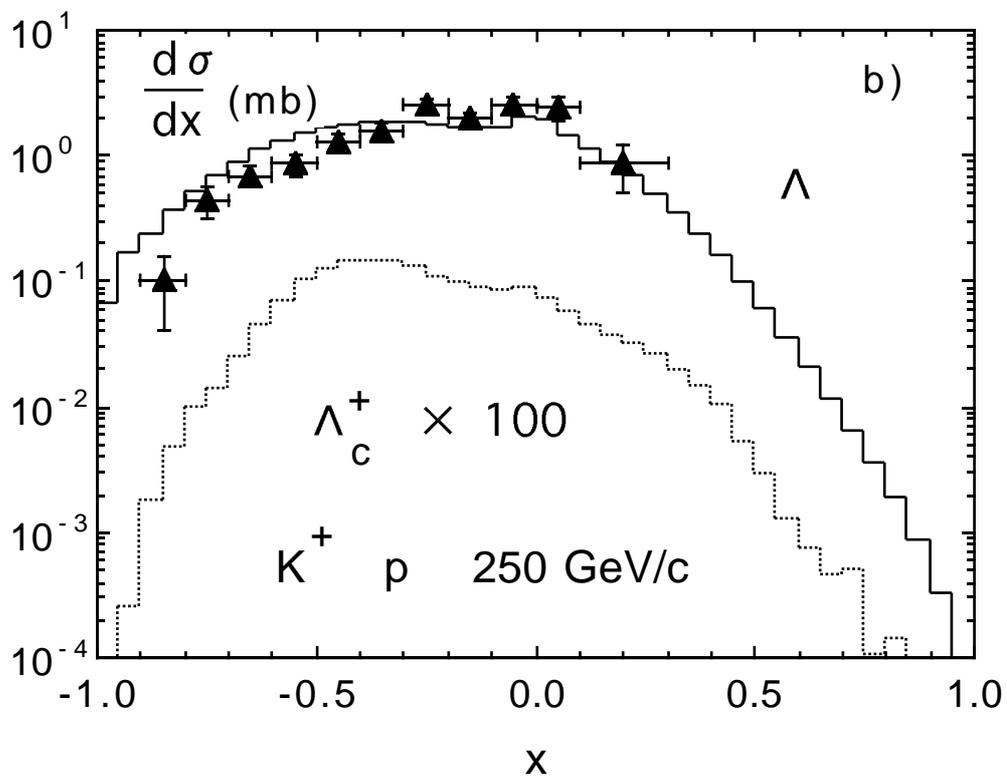
\begin{figure}
   \caption{Feynman $ x $ distributions of $ \Lambda $ and $ \Lambda_c $ in {\bf a} $ \pi^+p $ and {\bf b} $ K^+p $ collisions at $ p_L=250 $ GeV/c.  Experimental data are taken from [23].} 
   \label{fgr:pi&K to lamb}
\end{figure}
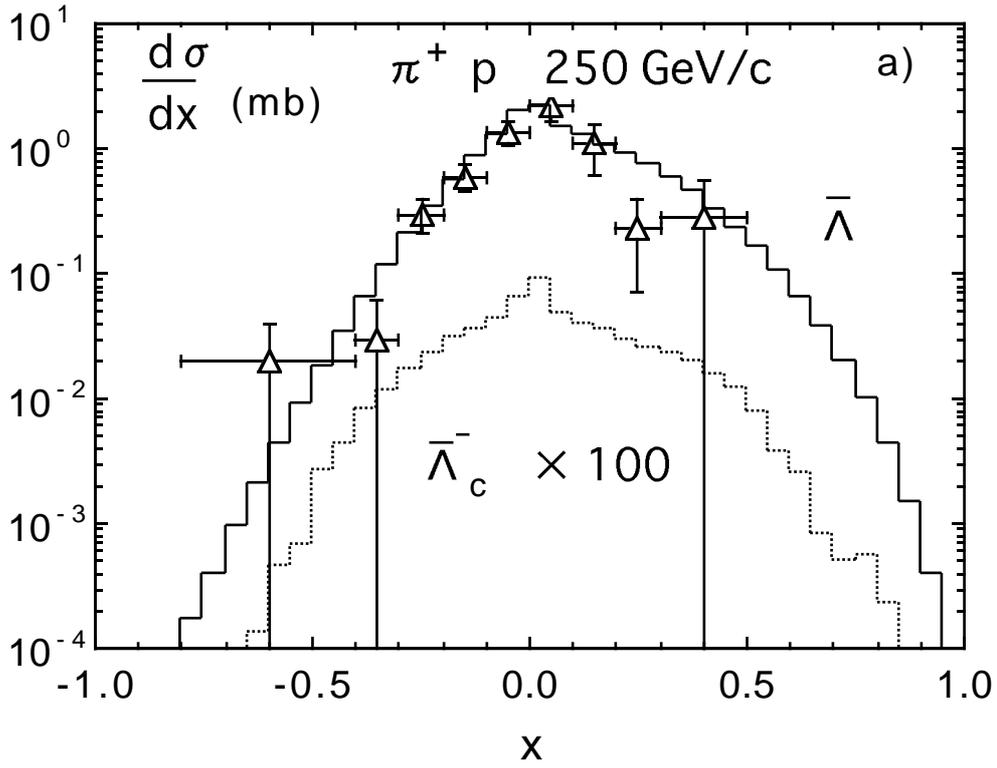
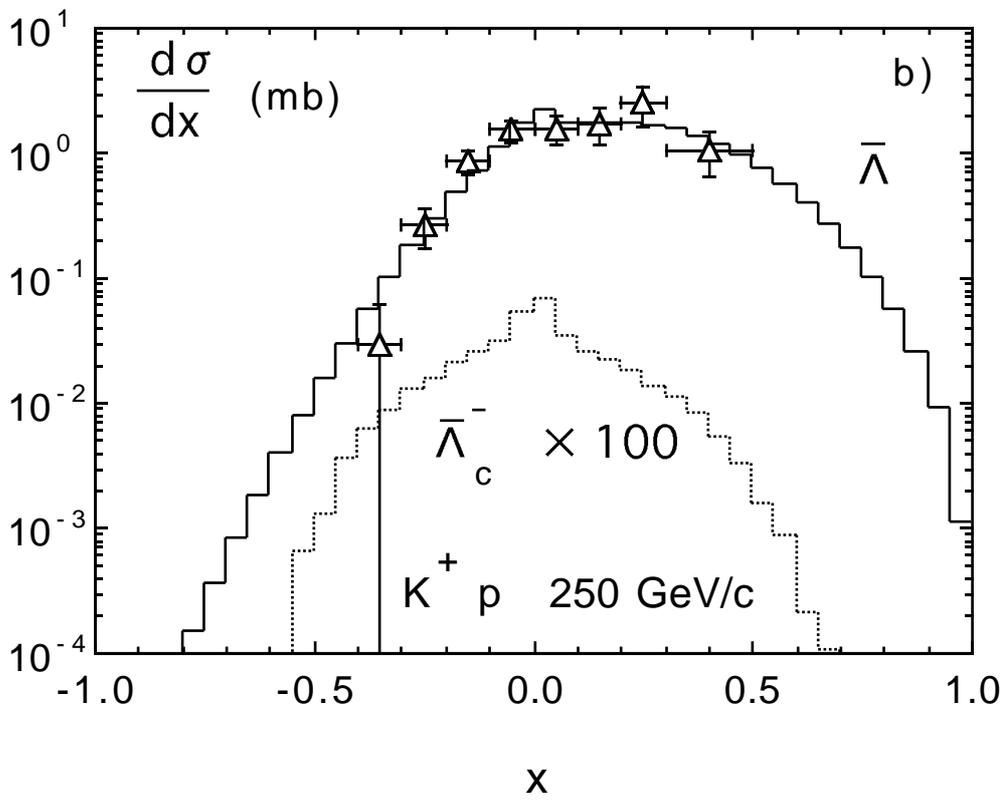
 \begin{figure}
   \caption{Feynman $ x $ distributions of $ \overline{\Lambda} $ and $ \overline{\Lambda_c} $ in {\bf a} $ \pi^+p $ and {\bf b} $ K^+p $ collisions at $ p_L=250 $ GeV/c.   Experimental data are taken from [23].}
   \label{fgr:pi&K to lamb_a}
\end{figure}


\begin{thebibliography}{6}

\bibitem{wa82}  M. Adamovich et al.(WA82): Phys. Lett. {\bf B305}(1993) 402
\bibitem{e769}  G. A. Alves et al.(E769): Phys. Rev. Lett. {\bf 72}(1994)  812
\bibitem{qs} J. Qiu and G. Sterman: Nucl. Phys. {\bf B353}(1991)  105,137
\bibitem{nde}  P. Nason, S. Dawson, K. Ellis: Nucl. Phys. {\bf B327}(1989) 49
\bibitem{PYTHIA}  H. U. Bengtsson, T. Sj\"{o}strand:
 Comput. Phys. Commun. {\bf 46} (1987)  43
\bibitem{vb}  R. Vogt, S. J. Brodsky: Nucl. Phys. {\bf B438}(1995)  261
\bibitem{Armesto} N. Armesto, C. Pajares, Yu. M. Shabelski: Preprint of Universidade de Santiago de Compostela, hep-ph/9506212, May 1995 
\bibitem{mnkt} H. Minakata: Phys. Rev. {\bf D20}(1979)  1656
\bibitem{cthkp} G. Cohen-Tannoudji,  A. El Hassouni,  J. Kalinowski,  R. Peschanski: Phys. Rev. {\bf D19}(1979)  3397
\bibitem{capella} A. Capella et al.: Phys. Lett. {\bf 81B}(1979) 68; A. Capella et al.: Z. Phys. {\bf C3}(1980) 329; A. Capella, U. P. Sukhatme, C.-I. Tan, J. Tran Thanh Van: Phys. Rep. {\bf 236}(1994)  225
\bibitem{andersson} B. Andersson, G. Gustafson, I. Holgersson, O. Mansson: Nucl. Phys. {\bf B178}(1981)  242;\\
B. Andersson, G. Gustafson, B. Nilsson-Almqvist: Nucl. Phys. {\bf B281}(1987)  289 

\bibitem{fi}  H. Fukuda, C.Iso: Prog. Theor. Phys. {\bf 57}(1977)  483,1663
\bibitem{innocente}  V. Innocente et al.: Phys. Lett. {\bf 169B}(1986)  285
\bibitem{ranft} J. Ranft: Z. Phys. {\bf C33}(1987)  517
\bibitem{qdq}  K. Kinoshita, H. Noda, T. Tashiro: Prog. Theor. Phys. {\bf 68}(1982)  1699, 2086;\\
 T. Tashiro, H. Noda, K. Kinoshita, C. Iso: Z. Phys. {\bf C35}(1987)  21 
\bibitem{qdqA} K. Kinoshita, H. Noda, T. Tashiro, J. Nagao: Int. Symp. on 
High Energy Nuclear Collisions and Quark Gluon Plasma, Kyoto 1991, 245
\bibitem{kkst}  T. Kanki, K. Kinoshita, H, Sumiyoshi, F. Takagi: Prog. Theor. Phys. Suppl. {\bf 97A}(1988) and {\bf 97B}(1989)  Chapters A, C, D, E and F.
\bibitem{km}  K. Kinoshita, A. Minaka: Prog. Theor. Phys. {\bf 81}(1989)  183
\bibitem{ab} M.Aguilar-Benites et al.: Z. Phys. {\bf C31}(1986) 491
\bibitem{Adamus} M.Adamus et al.(EHS-NA22): Z. Phys. {\bf C39}(1988) 311
\bibitem{Sheph} W.D.Shephard et al.: Phys. Rev. Lett. {\bf 27}(1972) 1164
\bibitem{Brenn} A.E.Brenner et al.: Phys. Rev. {\bf D26}(1982) 1497
\bibitem{na22baryon} I.V.Ajinenko et al.(EHS-NA22): Z. Phys. {\bf C44}(1989), 573 
\bibitem{na22} N.M.Agababyan et al.(EHS-NA22): Z. Phys. {\bf C46}(1990), 387  \\
 I.V.Ajinenko et al.(EHS-NA22): Z. Phys. {\bf C46}(1990), 525  \\
 N.M.Agababyan et al.(EHS-NA22): Z. Phys. {\bf C41}(1989), 539 

\end{thebibliography}
\end{document}